\documentclass[%
 aip,
 amsmath,amssymb,
 reprint,%
]{revtex4-1}

\usepackage{graphicx}% Include figure files
\usepackage{dcolumn}% Align table columns on decimal point
\usepackage{bm}% bold math
\usepackage[utf8]{inputenc}
\usepackage[T1]{fontenc}
\usepackage{mathptmx}
\usepackage{etoolbox}

\usepackage{hyperref}% add hypertext capabilities
\hypersetup{
    colorlinks=true,
    linkcolor=blue,
    filecolor=blue,      
    urlcolor=blue}

\makeatletter
\def\@email#1#2{%
 \endgroup
 \patchcmd{\titleblock@produce}
  {\frontmatter@RRAPformat}
  {\frontmatter@RRAPformat{\produce@RRAP{*#1\href{mailto:#2}{#2}}}\frontmatter@RRAPformat}
  {}{}
}%
\makeatother
\begin{document}

\preprint{AIP/123-QED}

\title{Electronic conduction in copper-graphene composites with functional impurities}
% Force line breaks with \\

\author{K. Nepal}
\email{kn478619@ohio.edu}
\altaffiliation{Department of Physics and Astronomy, Nanoscale and Quantum Phenomena Institute (NQPI), Ohio University}

\author{R. Hussein}
\altaffiliation{Department of Physics and Astronomy, Nanoscale and Quantum Phenomena Institute (NQPI), Ohio University}

\author{Y. Al-Majali}
\altaffiliation{Institute for Sustainable Energy and the Environment (ISEE), Ohio University}
\altaffiliation{Mechanical Engineering, Ohio University}

\author{D.A. Drabold}
\email{drabold@ohio.edu}
\altaffiliation{Department of Physics and Astronomy, Nanoscale and Quantum Phenomena Institute (NQPI), Ohio University}

% %\email{cu884120@ohio.edu}
% \ataffiliation[ME]{organization={Institute for Sustainable Energy and the Environment}, addressline ={Ohio University}, city = {Athens}, state={Ohio},postcode={ 45701}, country = {USA}}%
% \ataffiliation[ISEE]{organization={Mechanical Engineering}, addressline ={Ohio University}, city = {Athens}, state={Ohio},postcode={ 45701}, country = {USA}}%

% \author{D. A. Drabold}%

% \author{A. Author}
%  \altaffiliation[Also at ]{Physics Department, XYZ University.}%Lines break automatically or can be forced with \\
% \author{B. Author}%
%  \email{Second.Author@institution.edu.}
% \affiliation{ 
% Authors' institution and/or address%\\This line break forced with \textbackslash\textbackslash
% }%

% \author{C. Author}
%  \homepage{http://www.Second.institution.edu/~Charlie.Author.}
% \affiliation{%
% Second institution and/or address%\\This line break forced% with \\
% }%

\date{\today}% It is always \today, today,

\begin{abstract}
Coal-derived graphene-like material and its addition to FCC copper are investigated using \textit{ab initio} plane wave density functional theory (DFT). We explore ring disorder in the sp$^2$ carbon, and functional impurities such as oxides (-O), and hydroxides (-OH) that are common in coal-derived graphene. The electronic density of states analysis revealed localized states near the Fermi level, with functional groups contributing predominantly to states below the Fermi level, while carbon atoms in non hexagonal rings contributed mainly to states above it. The functionalization of graphene induces charge localization while ring disorder disrupts the continuous flow of electrons.  By projecting the electronic conductivity along specific spatial directions, we find that both the crystal orientation and the graphene purity significantly influence the anisotropy and magnitude of electronic transport in the composites. This study implicitly highlights the importance of structural stress to obtain improved electrical conductivity in such composites.
\end{abstract}

\maketitle

\section{INTRODUCTION}

Over the last decade, the development of ultra-conductive materials has become a focus in materials science, especially for applications in power transmission and motor systems. These materials, often synthesized under extreme thermo-physical conditions, have redefined the conventional understanding of the role of additives in metal matrices, leading to improvements in electrical performance \cite{Keerti_2022, MSE_2023, Gwalani_MD}. Copper has remained the main subject of this research due to its superior conductivity, while aluminum, being more abundant and lightweight, is gaining attention as a viable alternative. The introduction of graphene additives has shifted research away from traditional metal alloying methods toward advanced composite materials.\\

Coal-derived graphene has the potential to be an accessible and cost-effective source for high-performance ultra-conductors for a range of applications. Unlike conventional graphene production methods, which are expensive and resource-intensive \cite{exfoliate1, JH, JHreview, lin2019synthesis, hernandez2008high}, coal-derived graphene utilizes an abundant and inexpensive raw material, making large-scale production economically viable \cite{zhang2020structural, shi2021coal, yuan2021quantifying, xing2018preparation, zhou2012graphene, wang2022study}. Various methods exist for producing coal-derived graphene: direct carbonization through pyrolysis, chemical exfoliation involving oxidation-reduction, microwave-assisted methods utilizing rapid heating, and laser-induced graphene (LIG) offering precise control and patterned structures \cite{murakami2019fabrication, badenhorst2020assessment, liu2020pitch, zhou2012graphene, singh2023coal, liu2013repeated, jehad2020comparative, chamoli2017structural, ye2018laser, ye2019laser, duy2018laser}. This opens the door to widespread applications, from advanced electronics and energy storage to composite materials and high-efficiency conductors. As the world pushes toward greener, more efficient technologies, coal-derived graphene is poised to play a crucial role.  To ensure the effective industrial application of high-performance metal-graphene conductors, it is critical to systematically examine how graphene’s quality—specifically its purity, structure, and defect density—affects the composite’s electrical properties.\\

% Over the years, numerous studies have been reported that focus on a comprehensive understanding of ultraconducting materials. These studies have expanded from microscopic to macroscopic. 

This study, as an extension of a previous publication \cite{Nepal2025}, highlights the conduction activity between graphene and copper grains within composite microstructures at different interface orientations and provides additional evidence on the dependence of electronic conductivity on such conformations. We extend our analysis to more realistic copper-graphene composites by incorporating graphene with (i) topological disorder (amorphous graphene) at different defect densities and (ii) common-coal functionals -- oxides (-O) and/or hydroxides (-OH), akin to coal-derived graphene, into the material microstructures. Note that experimentally extruded composites undergo thermophysical processes, forming non-equilibrium / high-energy conformations. Our earlier study computed the dependence of electronic conductivity as a function of external pressure, showing that such high-energy conformation facilitates the efficient electronic interaction between metal matrices and graphene, leading to enhanced conductivity \cite{apl_2023,nepal_2024}.\\

We first describe the structural and electronic properties of sp$^2$ common coal. Next, we present a simulation of carbon-coal composites, aiming to provide insights for engineering optimized metal-graphene electronic devices. The Kubo-Greenwood formula (KGF) and space-projected conductivity (SPC) are used to spatially resolve electronic conductivity, demonstrating that both crystal orientation and graphene purity play critical roles in determining the electronic transport in the composites. We show that functional groups and ring disorder in the metal-graphene interface result in localized states near the Fermi level, with functional groups contributing primarily to states below it and disordered rings contributing to states above. These states act as carrier-trapping sites, disrupting coherent electronic interactions at the interface. Composites involving hydroxides showed higher electronic conductivity than oxide composites with pristine graphene; hydroxides are more or equally detrimental in composites with disordered graphene.  Next, we show that, with a high density of impurities, the electronic conductivity is largely unaffected by the interfacial purity of graphene, implying that impurity-driven scattering overshadows the benefits of using pristine graphene. A significant limitation of this paper is that only T = 0 K results are reported, thus neglecting phonon scattering. This will be discussed within an adiabatic approximation \cite{Abtew2007,Subedi_2022} in subsequent work.\\

The rest of this paper is organized as follows: Section \ref{Methodology} provides a summary of the computational models used in this work. Sections \ref{Method} and Section \ref{Theory} detail the simulation protocols and methodologies utilized for electronic properties and electronic conductivities calculations. In Section \ref{results}, the interface signatures and electronic transport of copper-carbon composites with functional impurities and ring disorder are reported.  Copper-composites based on crystalline graphene and functional impurities, and the coal-based metal composites (CMC) are detailed in section \ref{composites}.

\section{COMPUTATIONAL METHODOLOGY}\label{Methodology}

% \subsection{Models}\label{Models}

% \subsubsection{Graphene oxide}

% \subsubsection{Amorphous graphene and amorphous graphene oxide}

% \subsubsection{Interface structure of CMC}

% We simulated the CMC by constructing an interface model with a "sandwich" geometry shown in Figure~\ref{kfig1_models}. Here, we introduced the graphene layer to one of the low-energy crystallographic orientations, (111). Figure \ref{kfig1_models} is a schematic representation of the different geometries formed depicting the lattice mismatch between these metal orientations and aGr. Topologically disordered sp$^2$ carbon was placed in between two copper surfaces and relaxed forming an interface. The periodic boundary conditions ensure an infinite sheet of aGr between the copper matrix along the planar direction, while the graphene sheets are repeated for $\approx$ 20-25 \AA{} along the transverse direction. 
% \\

% Each interface model was relaxed under low-finite compressions. Table \ref{Table_models} lists the relaxed interface distance and the compression values associated with these composite models.  These compressions hope to mimic the external pressure exerted onto these composites during solid-phase extrusion \cite{Keerti_2022, MSE_2023, Gwalani_MD}. \\

\subsection{Electronic Structure Calculations}\label{Method}

The conjugate gradient algorithm, as implemented in VASP \cite{K17_vasp}, was applied for geometry optimization of the models. The convergence criterion for the energies on atoms of the configuration in an iteration step was set at $1 \times 10^{-6}$ eV. Geometric relaxation of layered graphene (amorphous and pristine), with and without functional impurities, employed $\Gamma$ point sampling of the Brillouin zone, while the Monkhorst-Pack scheme \cite{monkhorst1976special} with 2$\times$2$\times$1 \textbf{\textit{k}}-points was implemented for orthorhombic unit cell of the "sandwich" structure of metal-graphene composite. Periodic boundary conditions were used in all the calculations, such that an infinite sheet of planar graphene sheets is repeated for $\approx$ 20-25 \AA{} along the transverse direction. We used a plane-wave basis set with a kinetic energy cutoff of 420 eV.  Projected augmented wave (PAW) \cite{K18_paw} potentials to account for ion-electron interactions, and the generalized gradient approximation of Perdew-Burke-Ernzerhof (PBE) \cite{K19_pbe} as the exchange-correlation functional. We implemented Grimme’s van der Waals (D2) correction \cite{Grimme_D2} during structural relaxation of the interface models. A kinetic energy cut-off of 520 eV was implemented for electronic structure and electronic conductivity calculations.

% Geometric relaxation has been carried out using density-functional theory (DFT)-based VASP code \cite{K17_vasp}. We used a plane-wave basis set with a kinetic energy cutoff of 420 eV,  Projected augmented wave (PAW) \cite{K18_paw} potentials to account for ion-electron interactions, and the generalized gradient approximation of Perdew-Burke-Ernzerhof (PBE) \cite{K19_pbe} as the exchange-correlation functional. We implemented Grimme’s van der Waals (D2) correction \cite{Grimme_D2} during structural relaxation. 

% The electronic structure for graphene was calculated at the Gamma point, while for the orthorhombic unit cell of the "sandwich" structure of metal-graphene composite, the Brillouin zone was sampled using the Monkhorst-Pack scheme \cite{monkhorst1976special} with 2$\times$2$\times$1 \textbf{\textit{k}}-points. The  \\

\begin{figure*}
	\includegraphics[width=\textwidth]{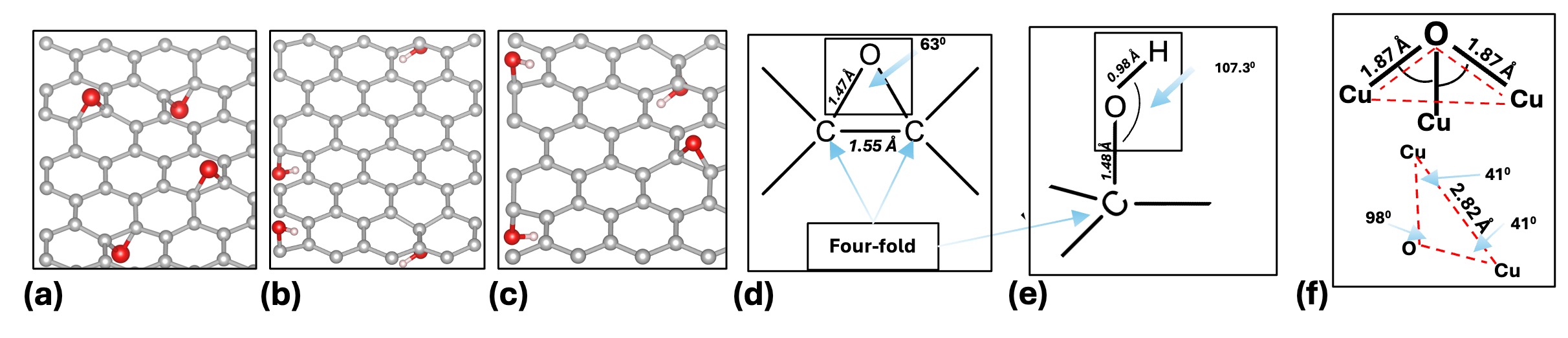}
	\caption{Pristine  sp$^2$ graphene with functional impurities (a) epoxides, (b) hydroxides, (c) epoxides and hydroxides. (d-e) local bonding chemistry of the epoxide group and the hydroxide group, respectively. (f) triangular pyramid arrangement of copper and oxygen at the copper-graphene interface. The numbers are average values. Color nomenclature: Brown-copper, gray-carbon, red-oxygen, and white-hydrogen.}
 \label{kfig_gr_structure}
\end{figure*}

%   \begin{figure}[b!]
%   \centering
%  \includegraphics[width=0.945\linewidth]{kfig_models.png}
%         \caption{(a) A typical configuration of the copper-graphene interface model. (b) [i, ii, iii] depict a top view of the interface showing the arrangement of carbon atoms and copper (111), (110), and (100) grains, respectively. Green and grey spheres represent copper (Cu), and carbon (C) atoms, respectively.}
% 	\label{kfig1_models}
% \end{figure} 

\subsection{Electronic Conductivity, Space-projected Conductivity }\label{Theory}

We computed electronic conductivity using the Kubo-Greenwood Formula (KGF) \cite{Greenwood_1958, KGF1, KGF2, KGF3}. For the atomistic local distribution of conductivity, we computed the space-projected conductivity (SPC), a spatial decomposition of KGF. The methodology for SPC is detailed elsewhere \cite{subedi_pssb}, but we provide a concise description in Appendix  \ref{Theory_KGF_SPC}. 

For this work, we focus on the DC limit, where the frequency of the Kubo field goes to zero ($\omega = 0$) in Equation \ref{kgf}. Next, in Equation \ref{kgf}, single-particle Kohn-Sham states $\psi_{i,\textbf{\textit{k}}}$ were computed from static calculations in VASP, the Fermi-Dirac distribution was computed with a 1000 K smearing temperature, and the $\delta$ function by a Gaussian distribution with a width $ k_BT$ (where $k_B$ and $T$ are the Boltzmann constant and smearing temperature). While the exact conductivity value from Equation \ref{kgf} depends on the choice of Gaussian width, the qualitative analysis remains largely unaffected, as demonstrated in \cite{Subedi_2022}. \\

\begin{table*}[!t]
\centering
\setlength{\tabcolsep}{6pt}
\renewcommand{\arraystretch}{1.4}
\begin{tabular}{|c|c|c|ccc|c|}
\hline
\textbf{Model} & \textbf{Functional Group} & \textbf{C–O–C} [$^\circ$] & \textbf{C–O} [\AA] & \textbf{C–C (4-fold)} [\AA] & \textbf{O–H} [\AA] & \textbf{C–O–H} [$^\circ$] \\ 
\hline
Gr–O         & –O       & 63       & 1.47       & 1.56         & –          & –         \\ \hline
Gr–OH        & –OH      & –        & 1.48       & –            & 0.98       & 107.3     \\ \hline
Gr–O–OH      & –O / –OH & 64 / –   & 1.46 / 1.48 & 1.55 / –     & – / 0.98   & – / 107.1 \\ \hline
aGr–O        & –O       & 62       & 1.45       & 1.51         & –          & –         \\ \hline
aGr–OH       & –OH      & –        & 1.47       & –            & 0.98       & 106.8     \\ \hline
aGr–O–OH     & –O / –OH & 64 / –   & 1.46 / 1.50 & 1.56 / –     & – / 0.98   & – / 107.1 \\ \hline
\end{tabular}
\caption{Average structural parameters of functionalized graphene with oxygen and hydroxyl groups. C--O--C: carbon-oxygen-carbon bond angle, C--O: carbon-oxygen bond distance, C--C: carbon-carbon bond distance, O--H: oxygen-hydrogen bond distance, C--O--H: carbon-oxygen-hydrogen bond distance.}
\label{tab1_bond_data}
\end{table*}

%  In pristine graphene models, the unit cell is slightly adjusted to account for a 3.6\% lattice mismatch with the copper (111) surface.For a more simplified analysis of the conduction active site, space-projected $N^2(\epsilon_f)$, by projecting the squared density of states over the Kohn-Sham states at the Fermi level ($\epsilon_f$) is computed. Space-projected $N^2(\epsilon_f)$  is given by:

% \begin{equation}\label{DoS5}
% \bar{\zeta}(\epsilon_f,\textbf{\textit{r}}) = \frac{1}{M^2} \sum_{i,j} \eta_{i.j}(\textbf{\textit{r}}) \delta (\epsilon_i - \epsilon_f) \delta (\epsilon_j - \epsilon_f)
% \end{equation}
% where, ${\eta_{ij}}(\textbf{\textit{r}})$ is a non-negative density function defined as:
% \begin{equation}\label{SPDOS}
%     \eta_{ij}(\textbf{\textit{r}}) = \beta_{ij}  |\psi_i(\textbf{\textit{r}})|^2|\psi_j(\textbf{\textit{r}})|^2 
% \end{equation}
% and $\beta_{ij} = 1 / \int|\psi_i(\textbf{\textit{r}})|^2|\psi_j(\textbf{\textit{r}})|^2d\textbf{\textit{r}}$, such that the space-integral of equation \ref{DoS5} gives $N^2(\epsilon_f)$.

% \begin{figure*}[!t]
% 	\includegraphics[width=\textwidth]{gr_func_chgden.jpg}
% 	\caption{Structure}
%  \label{fig_2}
% \end{figure*}

\begin{figure*}[!t]
	\includegraphics[width=\textwidth]{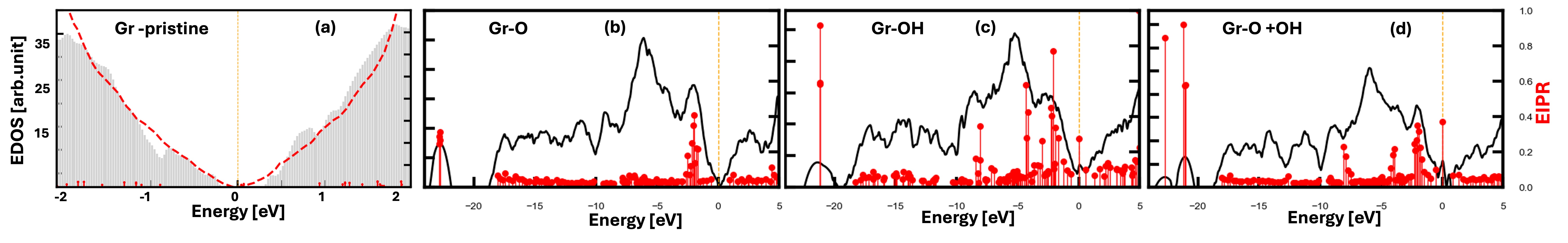}
	\caption{ [Left-axis]](a) Electronic density of states (EDOS) for pristine graphene. The red dashed line is EDOS from Reference \cite{Yu_CRC}, (b-d) Electronic Density of states for pristine graphene with functional impurities. [right-axis] Electronic Inverse Participation Ratio (EIPR) for the same models. Localized states are indicated by high EIPR values for graphene with functional impurities. The yellow vertical line represents the Fermi level.}
 \label{fig_3}
\end{figure*}
\begin{figure*}
	\includegraphics[width=\textwidth]{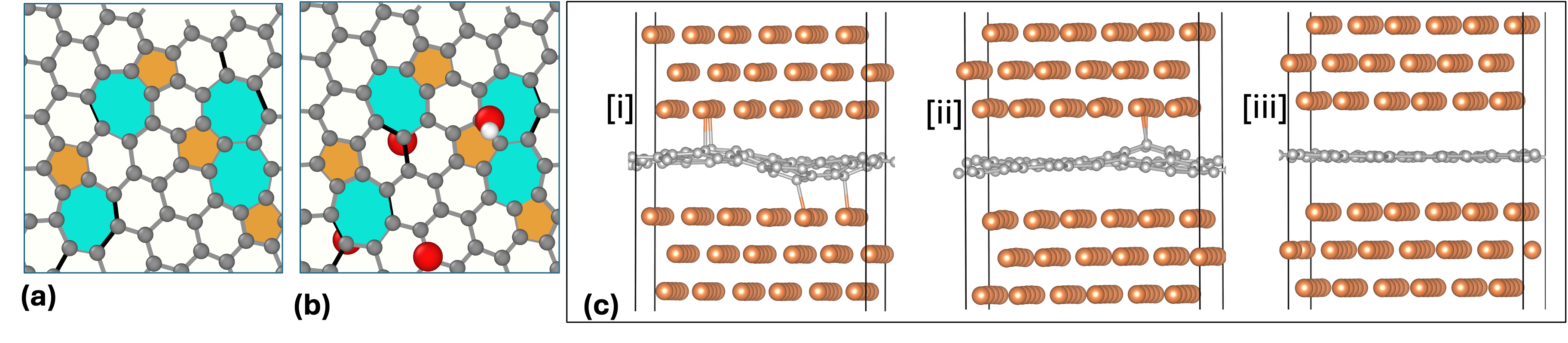}
	\caption{(a) Amorphous graphene showing sp$^2$ non-hexagonal rings. (b) Functional impurities (oxide and hydroxide groups) in amorphous graphene. Colored-rings: cyan-heptagon, yellow-pentagon. Black colored bonds correspond to long bond lengths in amorphous graphene. (c) Relaxed structure of amorphous graphene with increasing number of hexagonal carbon rings, respectively. Color nomenclature: Brown-copper, gray-carbon, red-oxygen, and white-hydrogen.} 
 \label{kfig_agr_structure}
\end{figure*}

\section{RESULTS AND DISCUSSIONS}\label{results}

\subsection{Structural and electronic properties of graphene with functional impurities}\label{gr_functional}
 Functional impurities (oxidation groups) and ring disorder (pentagons and heptagons at grain boundaries) are common graphene features.  Ring disorder may create local curvature (non-planarity) and influence the electronic properties of the material \cite{Yu_Puck, prl_raj}. In this section, we model and analyze the structure of graphene with topological disorder, followed by the effects of functional impurities, examining their corresponding electronic properties.

\subsubsection{Pristine graphene with functional 
impurities}\label{gr}
Oxides and hydroxides are common functional impurities in graphene. A layer of graphene with these functional impurities is created from a sp$^2$ carbon layer; a representative structure is shown in Figure \ref{kfig_gr_structure} (a). We simulated models with an epoxy group, the schematic of the relaxed layer is shown in Figure \ref{kfig_gr_structure}, models with hydroxide groups, and models with epoxy and hydroxide, see Figure \ref{kfig_gr_structure}(b-c). After energy minimization, significant structural modifications were observed in hexagonal graphene. The local bonding environment of graphene with different oxidation groups is shown in Figures \ref{kfig_gr_structure} (d-e). Epoxy group in graphene disrupts the sp$^2$ coordination of carbon atoms, forming fourfold-coordinated carbon atoms. This structural modification weakens the C-C bonds. These fourfold carbon atoms exhibit a longer bond length of $\approx$ 1.56 \AA{}. Longer bond lengths are reported for fourfold carbon atoms in graphene \cite{Odkhuu2013}, and $sp^3$ diamond \cite{Yi2022}. The average bond length between C and O was 1.47 \AA{}. In the model with hydroxide groups, the O-H bond length was observed at 0.98 \AA{}, while the C-O bond length was $\approx$ 1.48 \AA{}. A C--O--H bondangle was $\approx$ 107.3$^{\circ}$. In models with oxides and hydroxides, singly-bonded C and O atoms at 1.46 \AA{}, while double-bonded C and O showed a shorter bond length of $\approx$ 1.23 \AA{}. Refer to Table \ref{tab1_bond_data} for the summarized structural parameters for these different models.\\

The electronic density of states (EDOS) and the electronic inverse participation ratio (EIPR) of graphene with different oxidation groups were studied to analyze the influence of structural modification on electronic properties. The EDOS at the Fermi level ($E_f$) provides insight into the transport processes in the materials. With extended states at $E_{F}$, as illustrated in \cite{Nepal2025}, the conductivity scales as the square of the electronic density of states at the Fermi level [$\sigma \propto N^2(E_f)$], and regions of conduction activity in the material microstructure can be mapped out by spatial projection of $N^2(E_f)$ \cite{Nepal2025}. To enhance electronic conductivity, one needs to modify the material's microstructure to have an increased density of states at the Fermi level.

For qualitative comparison, first, the EDOS and EIPR were computed for pristine graphene, shown in Figure \ref{fig_3} (a), left and right axes, respectively. Notice a clear EDOS minimum at the Fermi level, and electronic states are extended as indicated by low EIPR values (red vertical drop lines). Next, for models with oxides, shown in Figure \ref{fig_3}(b), depicts a shoulder peak at -2 eV below the Fermi level. EIPR corresponding to this peak shows some degree of localization (shown by red vertical droplines with higher IPR values). Atomic projection of these states shows that the major contribution to these localized states arises from the oxygen atom in the model. A similar shoulder peak at -2 eV below the Fermi level is observed for a graphene model with hydroxide, shown in Figure \ref{fig_3}(c). In the Gr-OH model, energies ranging from -8 eV to -2 eV below the Fermi level show several localized electronic states, as indicated by higher EIPR values (red vertical drop lines). For the Gr-O-OH model, localized electronic states contributed by both oxides and hydroxides below the Fermi level were observed, shown in Figure \ref{fig_3} (d). 

\subsubsection{Amorphous graphene with ring disorder and  functional impurities}\label{agr}

We simulated three topologically disordered graphene (amorphous graphene (aGr)) samples with varying defect densities, following a protocol detailed in \cite{prl_raj}. We parameterize the ring disorder as $n_6/n$; where $n_6$ and $n$ are the numbers of hexagonal and non-hexagonal carbon rings in amorphous graphene. For $n_6/n = 3.0$, there is one non-hexagonal ring for every 3 hexagonal rings. Three models with $n_6/n = 1.3, \ 2.2$, and $3.0$ were simulated to study the electronic consequence of topological disorder in graphene. Refer to Figure \ref{kfig_agr_structure} (a) for a representative model corresponding to $n_6/n = 3.0$. Next, oxides and hydroxides were introduced into aGr with defect density $n_6/n = 3.0$. The structural features associated with functional impurities are summarized in Table \ref{tab1_bond_data}. The representative structure of aGr with oxidation groups is shown in Figure \ref{kfig_agr_structure} (b). We label the models of aGr with oxides and hydroxides as aGr-O and aGr-OH, respectively, and models with both oxidation groups as aGr-O-OH. \\

\begin{figure*}[!t]
	\includegraphics[width=\textwidth]{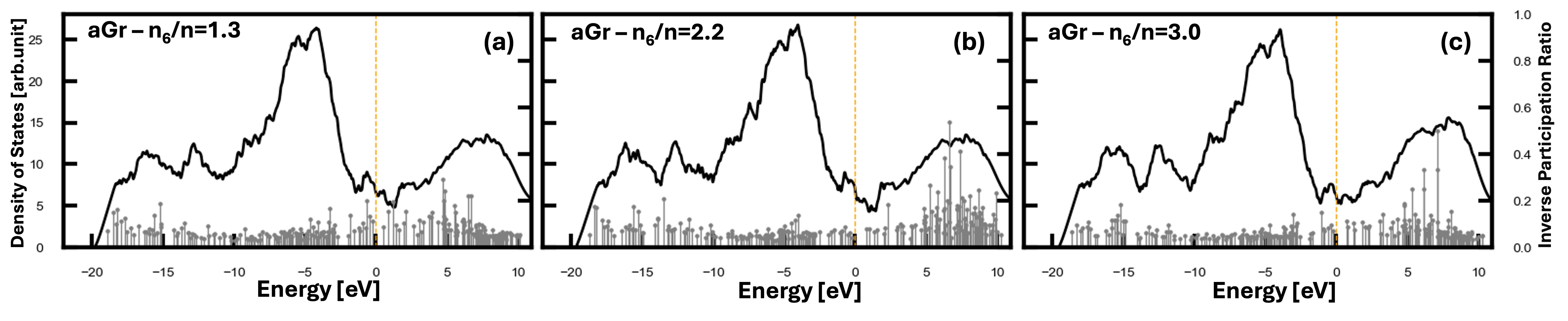}
	\caption{[Left axis] Electronic density of states and [Right axis] Electronic Inverse participation ratio for amorphous graphene models for varying disorder density (labeled in legends). Localized states are induced with topological disorder in graphene as indicated by high IPR values near 5 eV above the Fermi level. The yellow vertical line represents the Fermi level.}
 \label{fig_dos_agr_4}
\end{figure*}

 EDOS and EIPR for aGr with varying $n_6/n$ were computed and compared to pristine graphene, see Figure \ref{fig_dos_agr_4}. EDOS spectrum for aGr (black plots) models shows no distinct DoS minimum at the Fermi level, otherwise present in pristine Gr, shown in Figure \ref{fig_3} (a). In addition to that, electronic states at energies $\approx$ 5 eV above the Fermi level display some degree of localization, as indicated by their high IPR values (grey vertical drop lines). The projection of these states into atomic contributions showed that carbon atoms common to pentagons and heptagons contribute the most. While EDOS spectra in aGr with variations in the \( {n_6}/{n} \) ratio do not appear to cause significant changes in the EDOS spectra, we report new features in aGr (especially in the vicinity of the Fermi level) not present in pristine graphene.\\

\begin{figure*}[!t]
	\includegraphics[width=\textwidth]{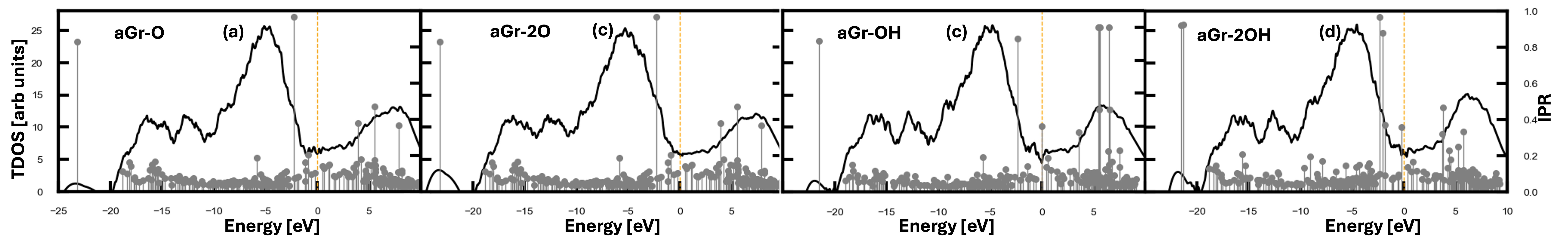}
	\caption{[Left-axis] Electronic density of states for amorphous graphene with varying functional impurities, as shown by plots' legends. [right-axis] Electronic Inverse Participation Ratio for the same models. (a-d) corresponds to amorphous graphene with one oxide, two oxides, one hydroxide and two hydroxides, respectively. Localized states are induced, indicated by high IPR values, with topological disorder [5 eV above the Fermi level] and functional impurities (the majority of localized states at energies below the Fermi level). The yellow vertical line represents the Fermi level. }
 \label{fig_dos_agr}
\end{figure*}

With oxides in aGr, localized states at $\approx$ 2 eV below the Fermi level were observed. These are illustrated in Figure \ref{fig_dos_agr}(a-b) for single oxide and two oxides in aGr, respectively. For aGr-OH, more states near the Fermi level are localized, indicated by higher IPR values in Figure \ref{fig_dos_agr}(c-d). In addition, localized electronic states $\approx$ 5 eV above the Fermi level in aGr indicate the presence of non-hexagonal rings in the material.

\begin{figure}
	\includegraphics[width=0.48\textwidth]{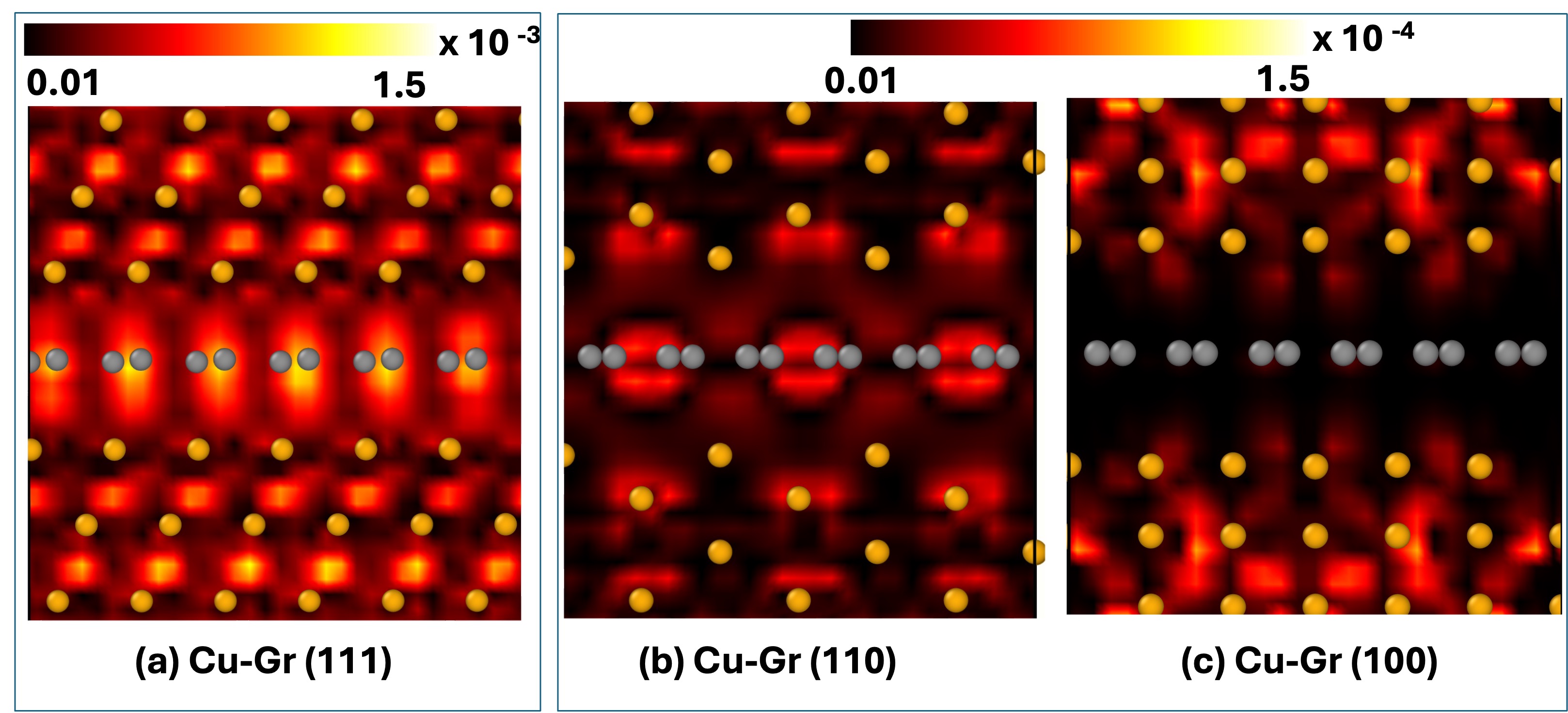}
	\caption{Space projected conductivity (in units of Siemens/cm/\AA{}$^3$) as a heat map showing the conduction active region at the copper-graphene interface. Figure (a-c) corresponds to the interface of graphene with (111), (110), and (100) copper grains, respectively. The colorbar shows the intensity of SPC increasing from dark to red to bright. Color Nomenclature: Brown: copper, grey: carbon.}
 \label{spc_orientation}
\end{figure}

\subsection{Copper-graphene composites}\label{composites}

\subsubsection{Orientation dependence of electronic conductivity}
Following the work by Nepal \textit{et. al.} \cite{Nepal2025}, strong orientational dependence in conductivity was observed in composites for different copper crystal orientations. Ideal low-energy interface structures were formed between the (111) and (110) copper surfaces and sp$^2$ graphene, exhibiting higher electronic conductivity \cite{Nepal2025}. To further analyze this, we computed the real-space projection of the KGF conductivity for composites with copper in three different crystallographic directions ((111), 110, and  (100)). \\

The connectivity and continuity of conduction activity within the composite are analyzed by computing a transverse SPC. By visualizing SPC through two-dimensional heatmap plots, we identify regions of high and low conductivity, aiding in the analysis of how different contacts between graphene and grain orientations impact the overall electronic performance of the composite. The "Top FCC"  registry formed between graphene and (111) copper induces a continuous network or path for carrier transport. This results in resonant mixing at the Fermi level, forming highly conductive pathways at the interface. This is highlighted by the high SPC regions in Figure \ref{spc_orientation} (a). (110) copper formed "bridge" registries with lower connectivity at the interface as compared to (111) (see Figure \ref{spc_orientation} (b), however, poor connectivity at the interface contact between graphene and (100) copper facilitates inferior carrier transport. Figure \ref{spc_orientation}(c) reveals this with low-intensity SPC values.  As noted in the previous literature, atomic alignments forming low-energy registry alleviate scattering, whereas misaligned geometries tend to act as sources of scattering \cite{nepal_2024, JPM_2024}. \\

%The correlation between the structural alignments at the interface and the electronic behavior might correspond to the observed pattern of the "Moire" in CVD-grown graphene in (110) and (111) copper substrates \cite{moire_1, moire_2}. \\
\begin{figure}[!b]
	\includegraphics[width=0.48\textwidth]{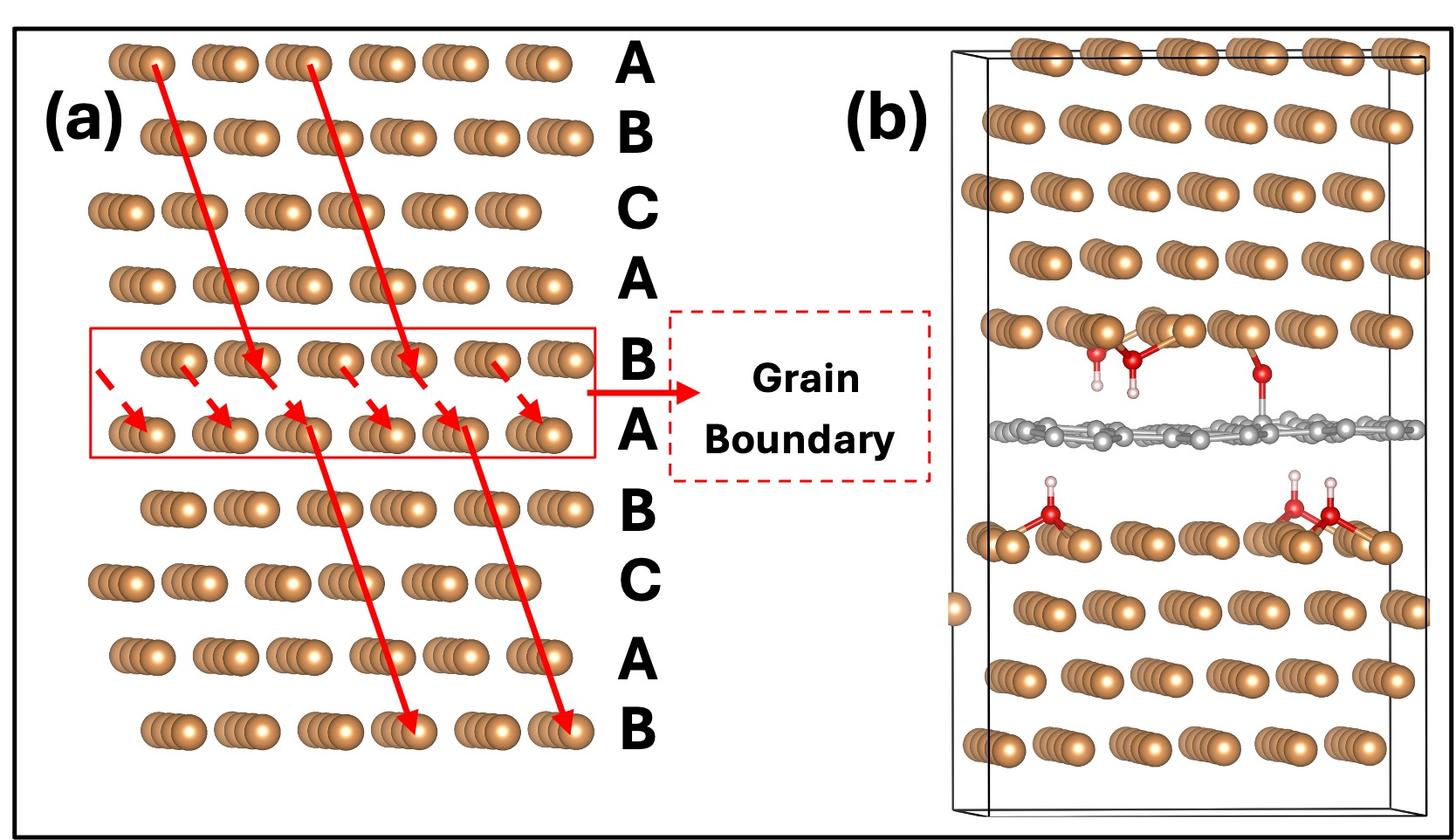}
	\caption{Relaxed structure of (a) (111) copper with a stacking fault forming a simple grain boundary model. The lines are guides to the eye to visualize the grain defect in the model. (b) interface model of amorphous graphene with functional impurities in (111) copper matrix. Color Nomenclature: Brown-copper, gray-carbon, red-oxygen, and white-hydrogen. }
 \label{kfig_gr_interface_cu}
\end{figure}

\begin{figure*}[!t]
	\includegraphics[width=\textwidth]{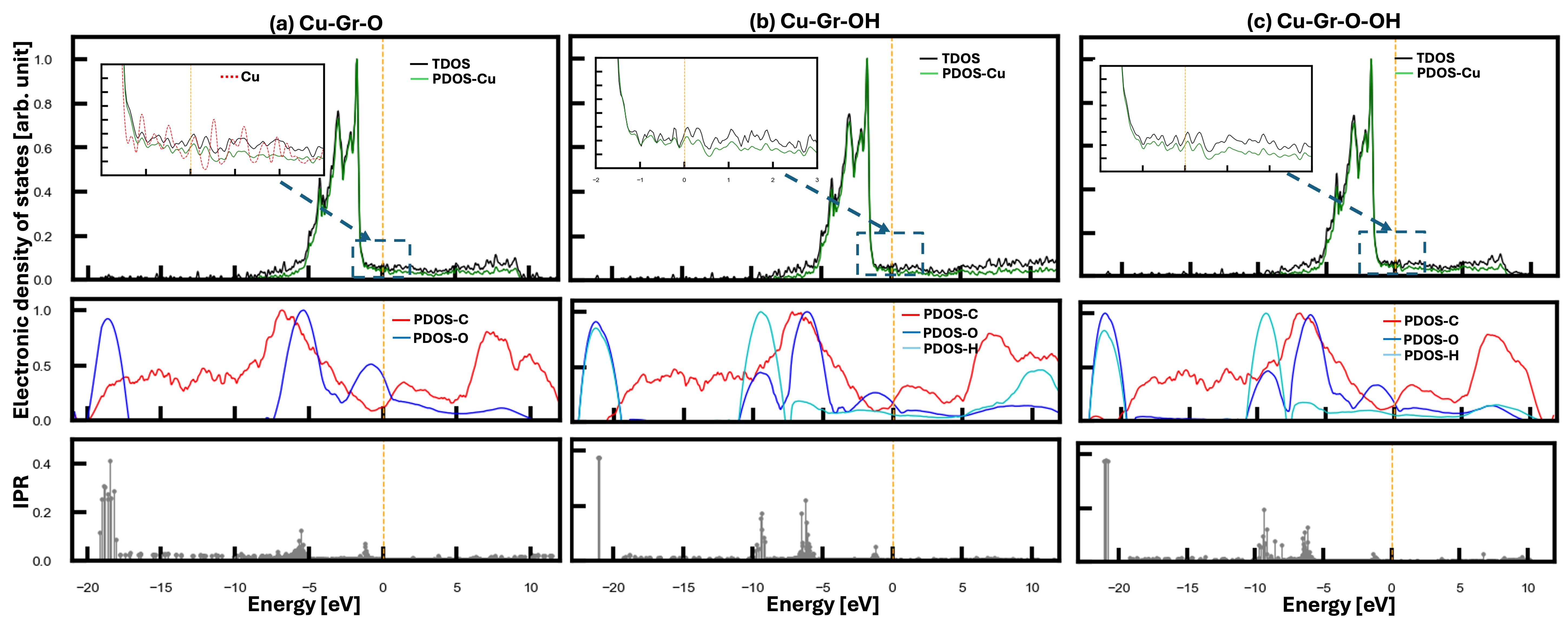}
	\caption{Total density of states (TDOS) in copper-graphene composites (Cu-Gr) with different functional groups shown in black plots. Projected density of states into copper (green plots), carbon atoms (red plots), oxygen atoms (blue plots ), and hydrogen (cyan plots). The electronic inverse participation ratio is shown as a grey colored vertical dropline. The Fermi level is shifted to zero, shown by yellow vertical lines in each plot. (a-c) respectively corresponds to copper-graphene composites with oxides, hydroxides, and both oxidation groups. The inset in (a) compares the PDOS of Cu and TDOS of the composites with the copper-only model (red dashed lines)  in the vicinity of the Fermi level. }
 \label{kfig_gr_cu_functional_imp}
\end{figure*}

For the rest of the paper, we work on the (111) oriented copper composites. An orthorhombic cell of 240 atoms of Cu-only system with a fault layer is modeled, shown in Figure \ref{kfig_gr_interface_cu}(a), representing a simple realistic grain boundary model, for comparison purposes. The KGF conductivity for the copper-only system was scaled to the standard IACS value ($\sigma_{Cu}$) $\approx$ 5.8  $\times $ $10^7$ Siemens/m. Whenever required, a comparison is made with the pressurized copper-graphene composite model, which exhibits an electronic conductivity $ \approx$ 118 \% IACS. We computed conductivities from Equation \ref{kgf} for smearing widths 0.05-0.005 eV that approximate a delta function. For larger supercells that have a relatively
dense electronic density of states near the Fermi level, a smaller broadening width can be invoked, which provides better agreement with experiment, as reported by \cite{Subedi_2022} for FCC aluminum. For this work,
we find that a smearing width of 0.005 eV is in close agreement with the experiment. The dependence of electronic conductivity on the smearing width and other factors such as the number of atoms, smearing temperature, and the energy cut-off for finite-size cells is discussed in the References \cite{parameter_1, parameter_2, parameter_3}. In addition, References \cite{apl_2023,nepal_2024} detail the dependence of electronic conductivity in aluminum and copper-based composites with external pressure.   \\

\subsubsection{Pristine Graphene with functional impurities in copper matrix}\label{grcomposites}

\paragraph{\textbf{Structural properties}}
The interface structures with the graphene structure containing common coal functional groups were created and structurally relaxed (representative structure for a relaxed interface is shown in Figure \ref{kfig_gr_interface_cu} (b)). Post-relaxation, the functional groups -O- and -OH migrate from their interstitial sites towards interface copper atoms (not all functional O migrate, but all -OH migrate). This results in a variety of bonding environments, forming a complicated structure at the metal-graphene interface. Hydroxides attached to carbon atoms migrate to form a metal hydroxide (Cu-OH), while the oxide group forms metal oxides (Cu-O-Cu).  Oxygen also formed an interlink connecting interface carbon and copper (forming a Cu-O-C structure). A more complicated tetrahedron geometry with Cu$_3$O chemical composition, shown in Figure \ref{kfig_gr_structure} (f) [top], was observed at the interface. Oxygen sits at the apex of three neighboring planar copper atoms such that any Cu--O--Cu forms a perfectly planar structure (sum of internal angles is equal to 180 degrees), as illustrated in Figure \ref{kfig_gr_structure} (f)[bottom]. These copper atoms form a distorted local configuration exhibiting longer Cu--Cu bond lengths ($d_{Cu-Cu} \ \approx$  2.82 \AA{}). This suggests that copper reacts strongly with oxygen, forming complex chemical environments at the interface. The microstructure of the extruded copper-graphene composites is likely to have regions of copper oxide, leading to localized oxidation at the interface. Our findings can guide an experiment that provides in-depth interfacial chemistry of the material and engineers the material for optimal efficiency. \\

%Table \ref{Table_Gr_Interface} summarizes the structural parameters of the interface structure.\\

%This is reminiscent of an "in situ" reduction during a metal-catalyzed reduction process in which graphene oxide (GO) undergoes a transformation into reduced graphene oxide (rGO) on a metal surface.

\paragraph{\textbf{Electronic density of states with -O and -OH}}

The total density of states (black plots) and the projected density of states in copper (green plots), carbon (red plots), oxygen (red plots), and hydrogen (cyan plots) atoms for copper-graphene composite with functional impurities are shown in Figure \ref{kfig_gr_cu_functional_imp}. Plots in the last row of Figure \ref{kfig_gr_cu_functional_imp} show the electronic inverse participation ratio as grey-colored drop lines. For all plots in Figures \ref{kfig_gr_cu_functional_imp}, yellow-colored vertical lines represent the Fermi level shifted to zero. Insets in the first row in Figure \ref{kfig_gr_cu_functional_imp} compare the total density of states to the projected density of states into copper atoms and the density of states for the Cu-only model (red-dashed line). The projected density of state at the Fermi level for copper atoms in the composite is comparable to the TDOS of the Cu-only system; meanwhile, this is slightly lower than the TDOS in the copper-graphene system. This suggests an increase in electronic states near the Fermi level with graphene additions, an electronic characteristic consistent with previous reports \cite{apl_2023,nepal_2024}. \\

For the Cu-Gr-O model, the PDOS of oxygen atoms shows small peaks at energies -1.0 and -6.0 eV below the Fermi level. For the Cu-Gr-OH model, the PDOS into oxygen and hydrogen atoms peaks at energies -1.0, -6.0, and -9.0 eV below the Fermi level. Note an additional peak at -9.0 eV for the Cu-Gr-OH model, dominantly from the H. Peaks at energies -1.0, -6.0, and -9.0 eV below the Fermi level for the Cu-Gr-O-OH models were observed. See Figures in the second row in Figure \ref{kfig_gr_cu_functional_imp}. For each composite model with functional impurities, the localized electronic states, as illustrated with the higher electronic inverse participation ratio, shown in the bottom row in Figure \ref{kfig_gr_cu_functional_imp} correspond to the energies where the PDOS into oxygen and hydrogen atoms peaks.\\

\begin{figure}[!t]
	\includegraphics[width=0.45\textwidth]{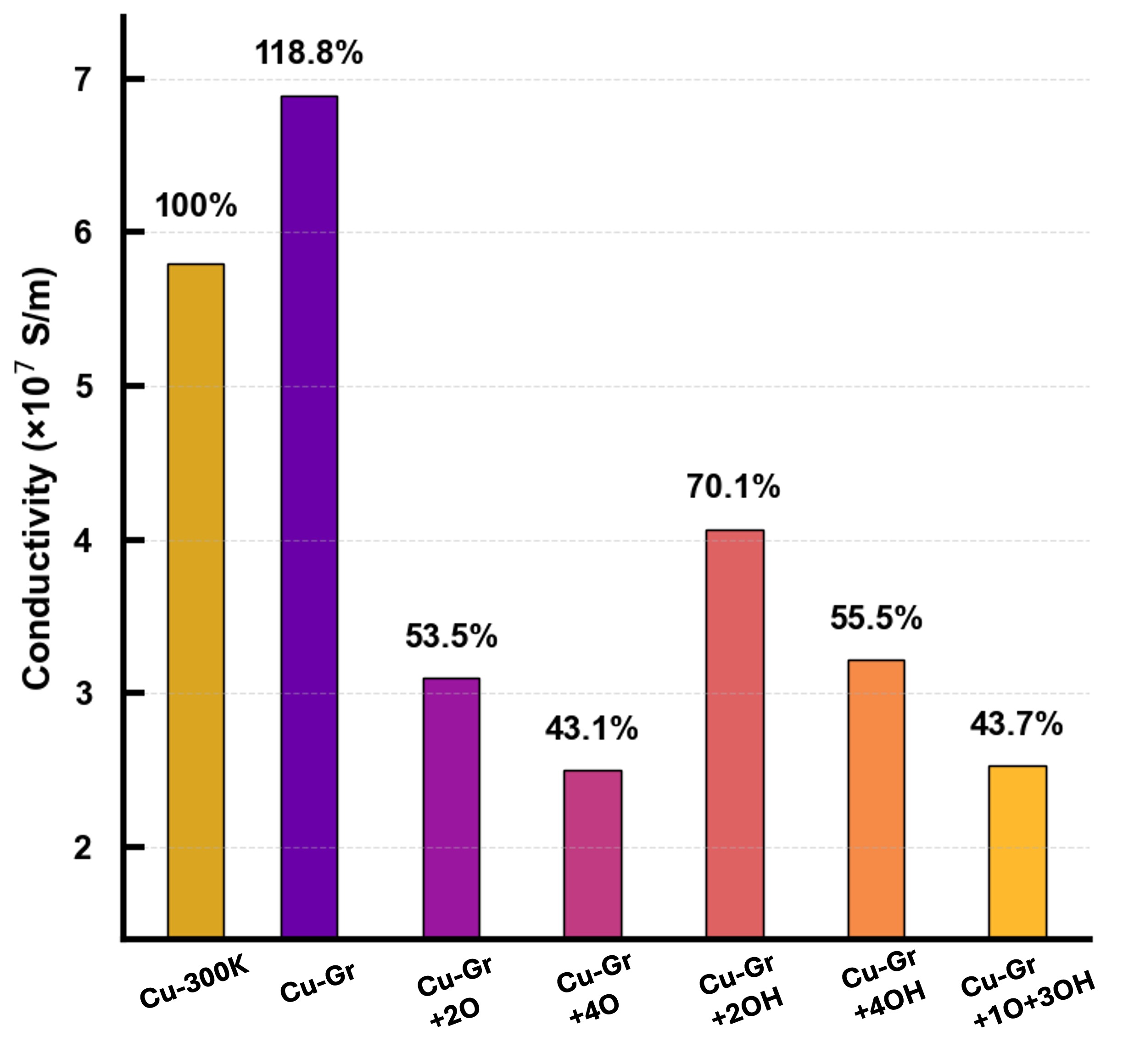}
	\caption{Electronic conductivity in Cu-Gr composites with oxides and hydroxides relative to the copper only model calculated at 300 K.}
 \label{kgf_bader_cu_gr_impurities}
\end{figure}

\paragraph{\textbf{Electronic conductivity with -O and -OH}}
Next, the conductivity of the interface models was computed and compared to the conductivity of a pure copper-only and impurity-free copper-graphene composite. Detailed electronic and transport behavior of impurity-free copper-graphene composites has been previously reported \cite{apl_2023,Nepal2025}. The EDOS shows that functional impurities have some degree of consequence for the localization of electronic states, intuitively expecting a significant effect on the transport properties of the composite. For the Cu-Gr interface with increasing oxidation groups, the electronic conductivity decreased. With two and four oxides, the material exhibits the electronic conductivity of $\approx \ 54 \% \ $  and $\approx \ 43 \% \ $ IACS. The average Bader charge gain by carbon atoms was ($\bigtriangleup q$) $\approx$ 0.02 electrons, while the charge accumulation to oxygen atoms at the interface is $\bigtriangleup q$ $\approx$ 0.9 electrons. This is a reduction of 40 \% of the Bader charge on carbon atoms from the interface copper as compared to the impurity-free Cu-Gr composite. With -OH functional impurities, the Bader charge gain in carbon atoms was 0.03 electrons, with significant charge accumulation in oxygen ( $\rangle $ 1 electrons), the majority of which is from hydrogen atoms. Composites with hydroxides exhibited slightly higher electronic conductivity as compared to oxides in the composite interface. Cu-Gr composite with two OH functional groups exhibits electronic conductivity $\approx$ 70.1 \% IACS, while this value reduces to 55.5 \% with four hydroxide groups. Figure \ref{kgf_bader_cu_gr_impurities} summarizes the conductivity results for composites with varying functional impurities compared to the copper system computed at 300K .\\

% factorGr = 6.9015

% polyCrystal (111-110)Cu = 18583.84
% Cu-Gr          = 20153.40
% Cu-Gr-KN = 26660.7
% Cu-Gr-Kashi = 24389.24
% factor  = 1.322888

% Cu-Gr-4O       = 7307.29 = 9666.73
% Cu-Gr-2O = 9072.31
% Cu-Gr-2OH = 11895.207
% Cu-Gr-3OH      = 9411.63 = 12450.53
% Cu-Gr-1O-3OH     =  7404.01 = 9794.67

% singleGrainBoundary Cu = 26626.09
% monocrystal Cu = 48390.59

% cu-agr1 = 9552.33
% cu-agr2 = 11016.18
% cu-agr3 =  12533.53

% cu-agr-1O = 9988.95
% cu-agr-2O = 8852.10
% cu-agr-4O = 7545.99
%Cu-aGr-1OH = 8568.77
% Cu-aGr- 2OH = 8435.09 
% cu-agr-2O-2OH = 7534.24

\begin{figure*}[!t]
	\includegraphics[width=\textwidth]{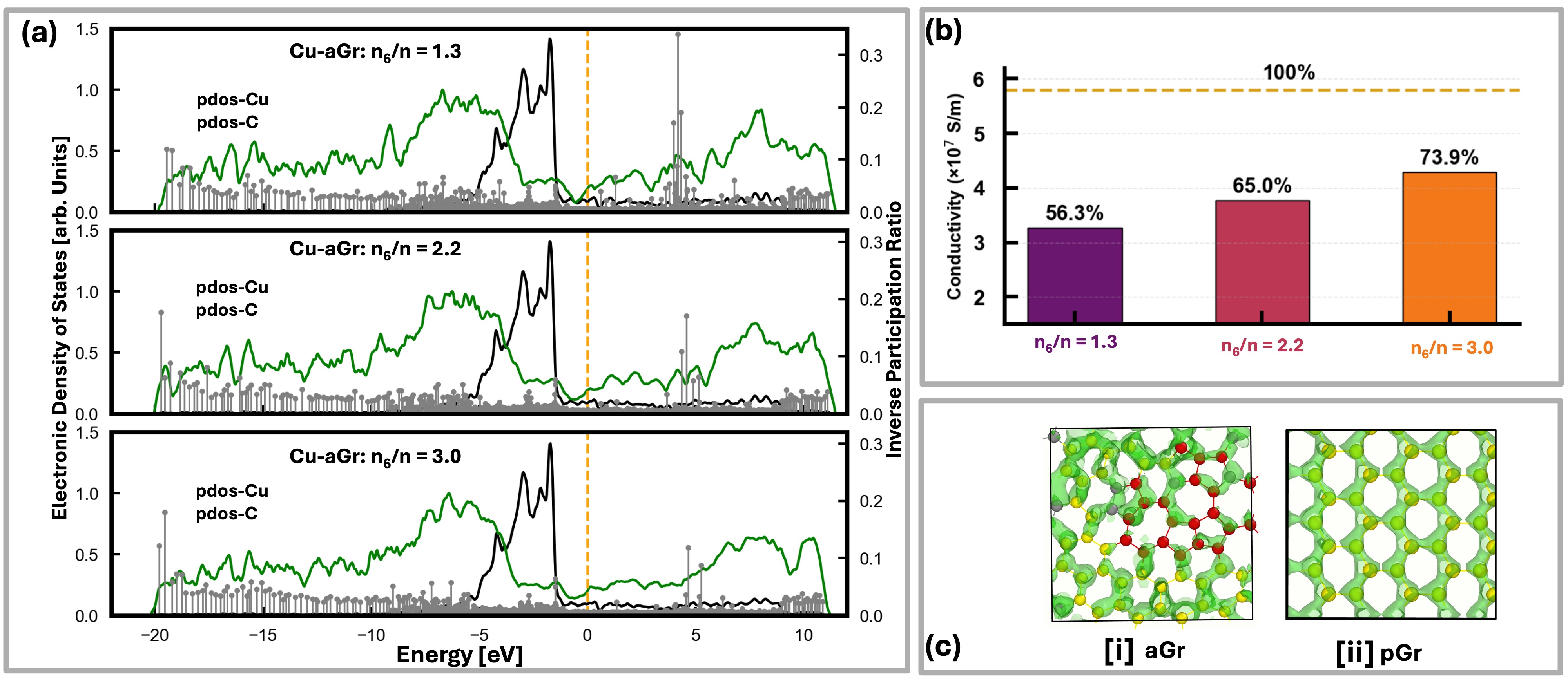}
	\caption{(a) Density of states projected into copper and carbon atoms for varying defect density in the graphene, as indicated by legends in subplots. (b) Relative KGF conductivities of Cu-aGr composites at varying $n_6/n$. (c) Space projected conductivity as an isosurface plot showing the electron conduction pathways in amorphous and pristine graphene. The red-colored sphere corresponds to carbon atoms participating in forming non-hexagonal rings.}
 \label{cu_agr_dos}
\end{figure*}

\begin{figure*}[!t]
\centering
	\includegraphics[width=\textwidth]{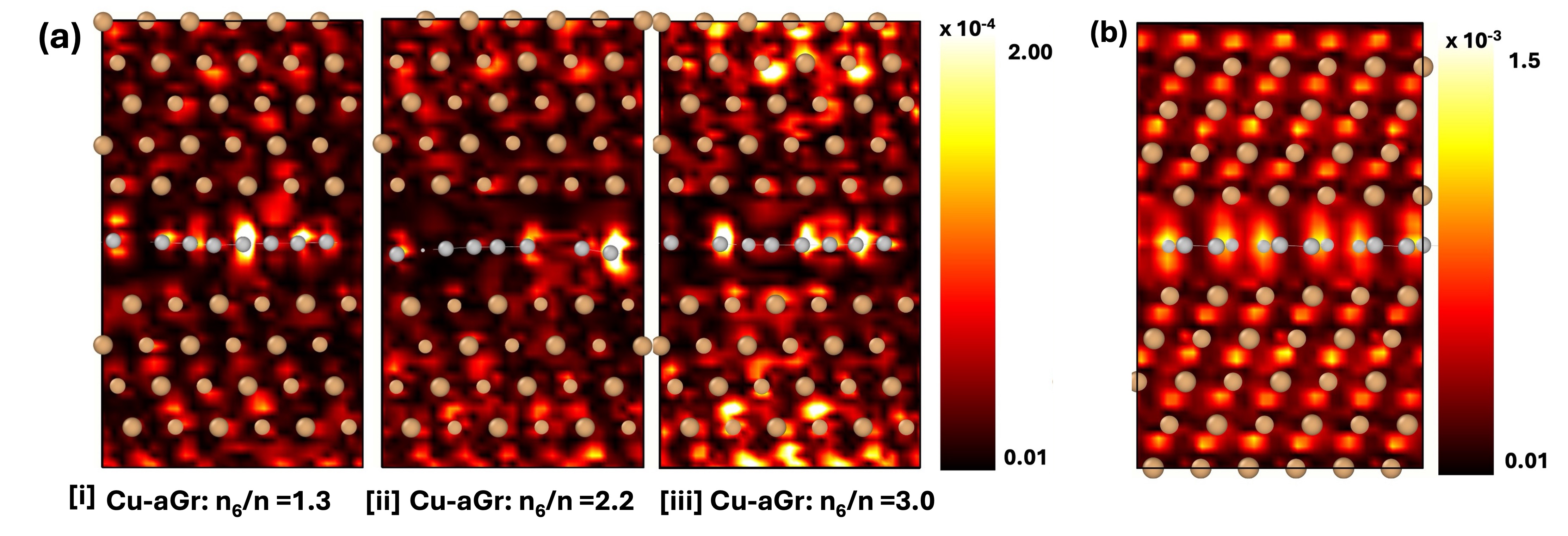}
	\caption{Transverse space projected conductivity (in units of Siemens/cm/\AA{}$^3$) as a heat map showing the conduction activity at copper-graphene microstructure with (a) amorphous graphene models with defect density $n_6/n $ = [i] 1.3 [ii] 2.2, and [iii] 3.0 and (b) pristine graphene. The colorbar shows the intensity of SPC increasing from dark to red to bright. Color Nomenclature: Brown: copper, grey: carbon.}
 \label{cu_agr_spc}
\end{figure*}
\begin{figure*}[!t]
	\includegraphics[width=\textwidth]{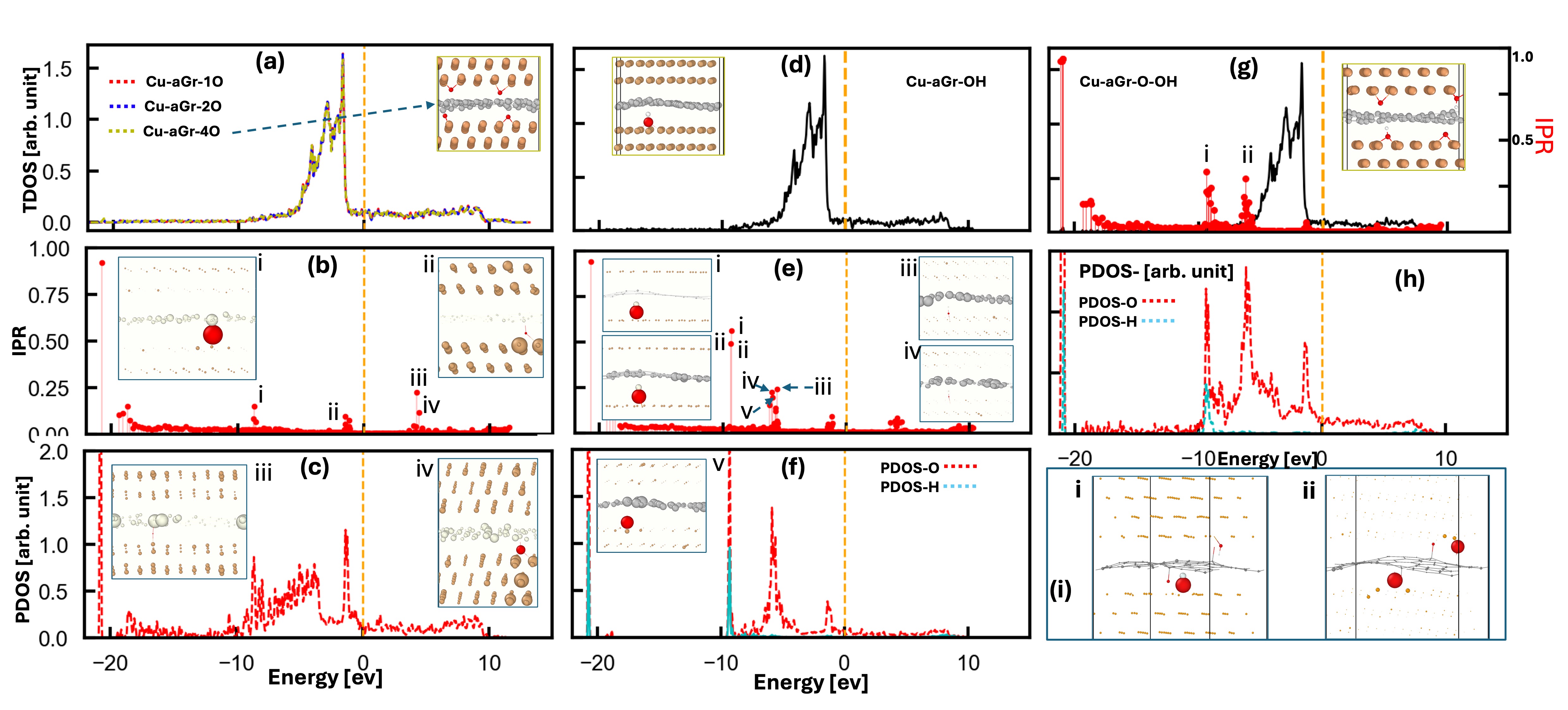}
	\caption{ (a) Total electronic density of states for copper-graphene composites with different oxide concentrations. The inset shows a representative interface structure of the relaxed composite model. (b) Electronic inverse participation ratio depicting some localized states at energies -22 eV, -8.5 eV, -1 eV, and + 5 eV relative to the Fermi level (shown by yellow vertical line at 0 eV), some labeled "i-iv". (c) Projected density of states into oxygen atoms. The peaks of PDOS into O atoms correspond to energies with localized states in (b), except at 5 eV above the Fermi level. Insets in (b) and (c) show the atom projection of these localized states, detailed in the text. (d) Total electronic density of states for copper-graphene composites with hydroxide. The inset shows a representative interface structure of the relaxed composite model with hydroxide in the interface. (e) Electronic inverse participation ratio depicting some localized states at energies -21 eV, -9 eV, -6 eV, and + 5 eV relative to the Fermi level (shown by yellow vertical line at 0 eV), some labeled "i-v". (f) Projected density of states into oxygen atoms (red plots) and hydrogen atoms (blue plots). Note the peaks at PDOS into O and H atoms correspond to energies with localized states in (e), except at 5 eV above the Fermi level. Insets in (e) and (f) show the atom projection of these localized states, detailed in the text. (g) [left axis] Total electronic density of states for copper-graphene composites with different oxide and hydroxide. The inset shows a representative interface structure of the relaxed composite model. [Right axis] Electronic inverse participation ratio depicting some localized states at energies -21 eV, -9 eV, -6 eV relative to the Fermi level (shown by yellow vertical line at 0 eV), some are labeled "i-ii". (h) Projected density of states into oxygen atoms (red plots) and hydrogen atoms (blue plots). Note the peaks at PDOS into O and H atoms correspond to energies with localized states in (g). (i) Atom projection of localized states labeled "i-ii" in (g), detailed in the text. }
 \label{cu_agr_dos_func}
\end{figure*}

\begin{figure}
	\includegraphics[width=0.5\textwidth]{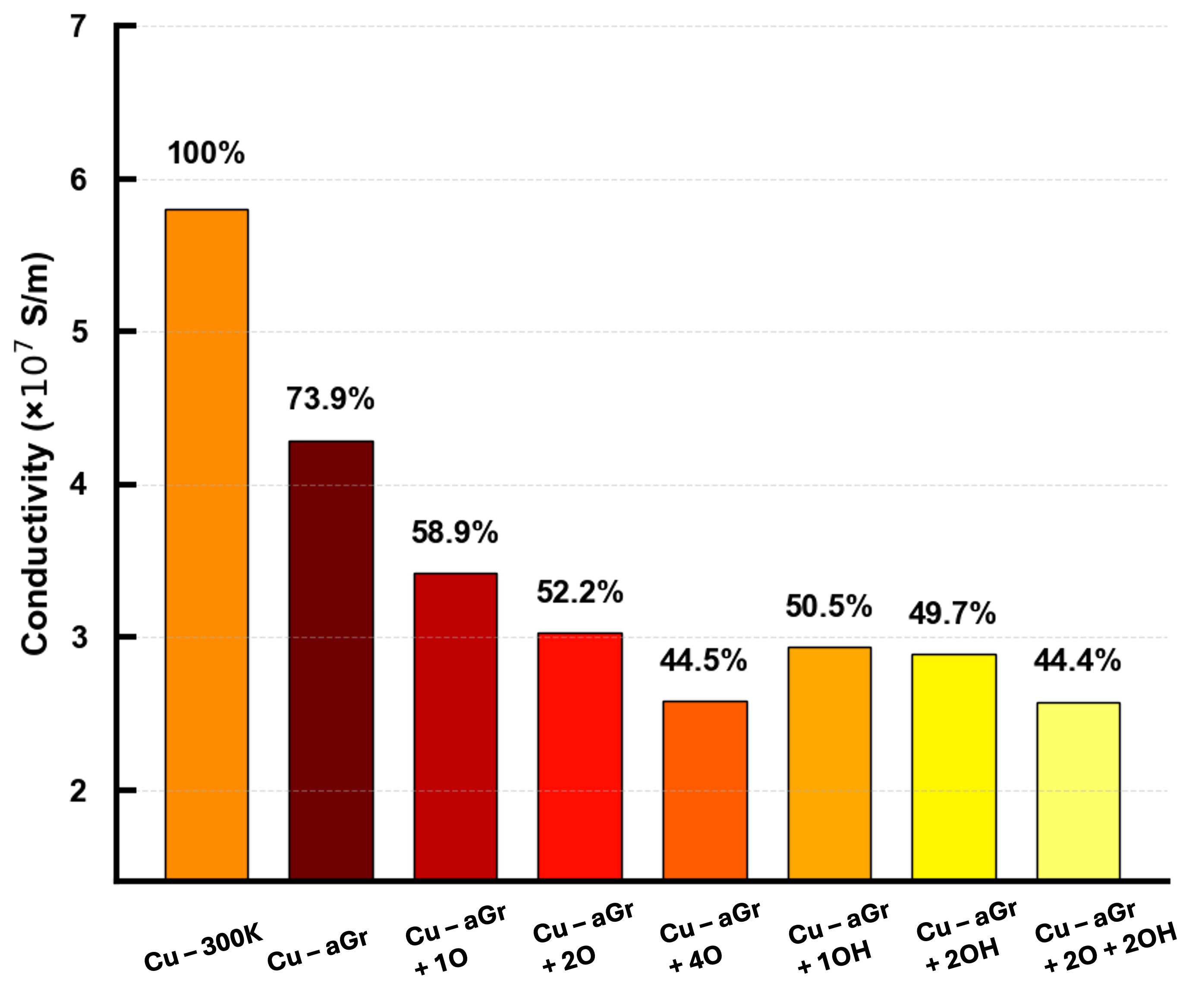}
	\caption{Relative KGF conductivities for copper-amorphous graphene composites with functional impurities compared to the standard IACS value for copper.  }
 \label{cu_agr_sigma_funct}
\end{figure}
\begin{figure*}
	\includegraphics[width=\textwidth]{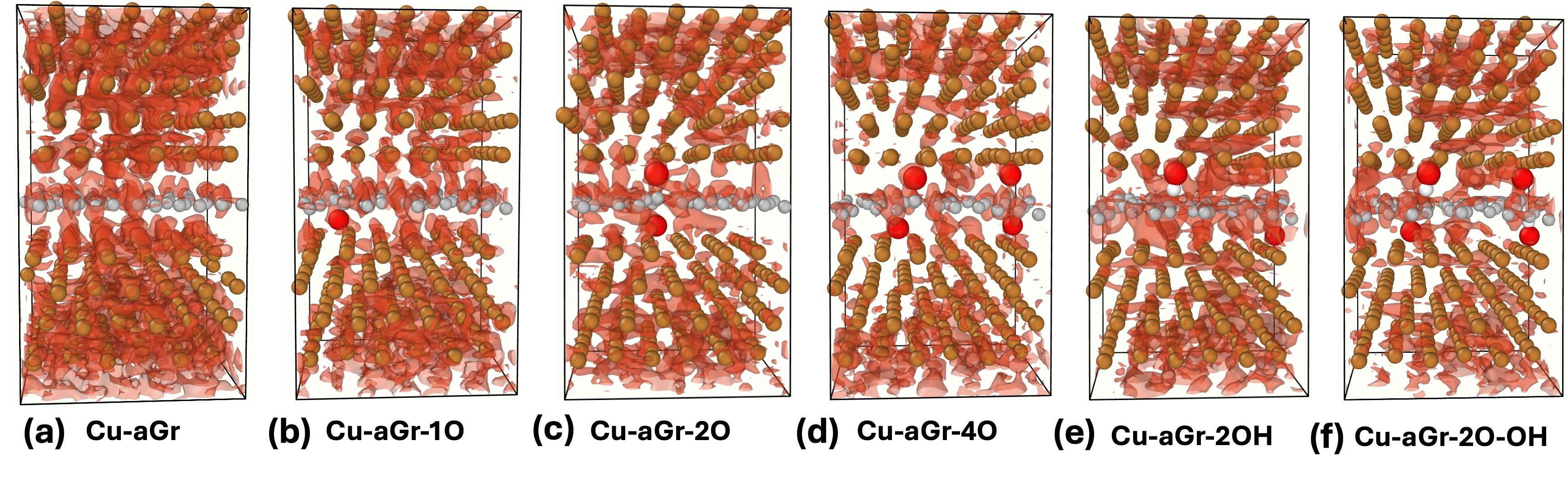}
	\caption{Space projected conductivity (in units of Siemens/cm/\AA{}$^3$) as an iso-surface plot showing the regions of active conduction at copper-amorphous graphene microstructure with functional impurities. We include 20 \% of the highest local contributions to SPC in each plot. Color Nomenclature: Brown-copper, gray-carbon, red-oxygen, and white-hydrogen. }
 \label{agr_metal_funct}
\end{figure*}
\subsubsection{Amorphous graphene in copper matrix }\label{agrcomposites}

Interface structures were created with amorphous graphene with a varying $n_6/n$. The amorphous graphene in the copper matrix forms an undulating structure, shown in Figure \ref{kfig_agr_structure} (c)[i-ii], due to nonhexagonal rings that cause puckering of the amorphous graphene \cite{Yu_Puck}. With increasing topological order, a planar structure is observed at the interface, highlighted in Figure \ref{kfig_agr_structure}(c)[iii]. This is an intrinsic nature of graphene - to form a planar structure at the interface, supported by the fact that pristine graphene forms a planar structure in a copper matrix, regardless of the number of graphene layers in well-defined stacking. The PDOS and IPR plots for Cu-aGr composites for varying $n_6/n$ show reduction of the localization of states (at energies 5 eV above the Fermi level), as highlighted with electronic states with low IPR values in Figure \ref{cu_agr_dos} (a), while the DOS spectrum shows no significant variations. The KGF conductivity of the composites varied with $n_6/n$ aGr in microstructure, providing insights into the sensitivity of electronic conductivity to topological disorder in graphene. The electronic conductivity for $n_6/n$ = 3.0 is ($\approx$ 74 \% IACS, with further reduction with lower $n_6/n$. Conductivities for $n_6/n$ = 2.2 and $n_6/n$ = 1.3 are $\approx$ 65 \% and $\approx$ 56 \% IACS. The comparison is illustrated in Figure \ref{cu_agr_dos} (b). Note that the electronic conductivity for Cu-aGr composite with $n_6/n$ = 3.0 is slightly higher than Cu-Gr composite with two OH functional groups in the microstructure. \\

Topological disorder in graphene affects the flow of carriers along the sp$^2$ connected carbon rings and also affects the overlap between carbon's $\pi$-orbital with the interface copper. To visualize this, space-projected conductivity along the plane of graphene in the composite's microstructure is computed and shown in Figure \ref{cu_agr_dos} (c)[i]. Non-hexagonal rings with connecting atoms (depicted by red-colored atoms) limit the continuous flow of electrons as opposed to a homogenous and continuous flow of electrons in pristine graphene, shown as an isosurface (green color) plot in Figure \ref{cu_agr_dos}(c) [ii]. By computing transverse space-projected conductivity, the electronic dynamics between the metal surface and pi orbitals in sp$^2$ graphene were assessed and compared to that of Cu-Gr composite. While there is poor electron activity at the interface for Cu-aGr composites, an increase at the interface and in the bulk was observed with reduced topological disorder. Figures \ref{cu_agr_spc} (a) [i--iii] show the SPC as a heat map corresponding to Cu-aGr interfaces with $n_6/n$ = 1.3, 2.2, and 3.0, respectively, and compared to the pristine Cu-Gr composite, see Figure \ref{cu_agr_spc}(b). The SPC intensity increases from black to red to bright, depicted by color bars.\\

Next, we introduce aGr with functional groups (-OH and -O) to model coal-derived graphene into a metal matrix. Coal-derived graphene is characterized by the presence of topological disorders and different oxidation groups on the graphene surface. For the rest of the analysis, we use aGr with $n_6/n$ = 3.0. After relaxation, migration of functional groups from graphene to the copper interface was observed.\\

%Table \ref{table3} summarizes the structural parameters of Cu-aGr composites with functional impurities.

\paragraph{\textbf{Cu-aGr with oxides}}
Electronic density of states for the composite with oxides is shown in Figure \ref{cu_agr_dos_func} (a). No drastic variation in DoS was observed with increasing oxide in the material. We observed regions in the energy spectrum with several electronic states exhibiting some extent of localization as indicated by high IPR values in Figure \ref{cu_agr_dos_func}(b). Localized states at energy -22 eV below the Fermi level are due to the oxide in the material microstructure.  We observed peaks at $\approx$-8.5 eV below the Fermi level (labeled "i"), -2 eV (labeled "ii"), and states "iii-iv" at energies $ \approx$ 5 eV above the Fermi level. By computing the projected DOS into the O atom, we see that oxygen present in the microstructure contributes to these localized states, shown in Figure \ref{cu_agr_dos_func} (c). However, O has no significant contribution in the localization of states above the Fermi level. The states i-iv are projected into the atoms weighted by their contribution to those states, shown as an inset in Figure \ref{cu_agr_dos_func}(b-c). The size of the sphere corresponds to the weight of atoms in the electronic state.  The state "i" is primarily localized in the O atom and a C atom in its neighborhood, while state "ii" is primarily localized in the interface copper atoms. Both states "iii" and "iv" are consequences of non-hexagonal rings in graphene, with state "iv" showing the mixed contribution from oxygen and copper atoms.

\paragraph{\textbf{Cu-aGr with hydroxide}}
Electronic density of states for the composite with -OH is shown in Figure \ref{cu_agr_dos_func} (d) with an inset showing a relaxed microstructure of Cu-aGr interface with the OH group. We observed regions in the energy spectrum with several electronic states exhibiting some extent of localization as indicated by high IPR values in Figure \ref{cu_agr_dos_func}(e). Notably, we observe two localized states (labeled "i" and "ii")  at $\approx$ -9.5 eV and a few states (labelled "i-iv") below -5 eV below the Fermi level. States "i-ii" are explicitly due to the -OH group present in the material microstructure, as shown by PDOS in H atoms in Figure \ref{cu_agr_dos_func} (f), and atom projected IPR weighted by their contribution, shown in Figure \ref{cu_agr_dos_func}(e) (inset). States "iii-v" has contribution from C and O atom, as shown by the insets in Figure \ref{cu_agr_dos_func} (e-f). \\

\paragraph{\textbf{Cu-aGr with oxides and hydroxides}}
Electronic density of states for the composite with -O and -OH is shown in Figure \ref{cu_agr_dos_func} (g) with an inset showing a relaxed microstructure of Cu-aGr interface. We observed regions in the energy spectrum with several electronic states exhibiting some extent of localization as indicated by high IPR values in Figure \ref{cu_agr_dos_func}(g) [right axis]. We observe states at $\approx$ -9.5 eV and -6 eV below the Fermi level. exhibiting localization. These states, labeled "i" and "ii", are explicitly due to the -OH and -O present in the system, as shown by PDOS plots in Figure \ref{cu_agr_dos_func} (h). Further analysis of these states was performed by projecting these states onto atoms, weighted by their contribution, shown in Figure \ref{cu_agr_dos_func}(i). \\

\paragraph{\textbf{Electronic Conductivity of Cu-aGr with -O and -OH}}
Functional impurities in the material microstructure resulted in the localization of electronic states. These atoms act as a site for trapping carriers, limiting the flow of electrons in the microstructure. This results in reduced electronic conductivity of the material. With increasing O groups in the material, the conductivity decreases linearly. The computed electronic conductivity for one, two, and four -O groups is $\approx$ 59 \%, $\approx$ 52 \%, and $\approx$ 45 \% IACS, respectively. The conductivity of the composite model with one and two -OH groups is $\approx$ 51 \% and 50 \% IACS. The Cu-aGr model with two -O and two -OH composite is $\approx$ 44 \% IACS. Figure \ref{cu_agr_sigma_funct} summarizes the KGF electronic conductivity for Cu-aGr models with different functional impurities. These observations show that the electronic conductivity of metal-graphene composites with a high impurity density likely remains similar regardless of the purity of graphene at the interface.  \\

The space-projected conductivity was calculated for these composite models and is shown as isosurface plots in Figure \ref{agr_metal_funct}. The reduction in total electronic conductivity with the introduction of functional impurities in the metal microstructure is due to diminished local SPC contributions, particularly evident by the weakened continuity of the isosurfaces at the interface. For aGr in metal matrix, the isosurfaces show partial connectivity across the graphene layers, and the connectivity recovers as we go to the bulk metal, see Figure \ref{agr_metal_funct} (a). However, with functional impurities in the microstructure, a breakdown in isosurface connectivity is seen, not only at the metal-graphene interfaces but also within the bulk metal, as shown in Figure \ref{agr_metal_funct}(b-f), indicating impaired electron transport in the metal microstructure.

% \subsection{High-temperature performance of metal-carbon composites}
% \begin{figure}
% 	\includegraphics[width=0.45\textwidth]{temp_copper.jpg}
% 	\caption{Log plot of the KGF resistivity for a cubic-256 atom of FCC Aluminum compared to the Bloch formula and experimental values computed from Reference \cite{} }
%  \label{fig_1}
% \end{figure}
\section*{Conclusion}

In conclusion, our study investigated the structural, electronic, and transport properties of copper-coal composites, detailing the role of copper orientation, graphene structure, defect density, and chemical impurities in the composites microstructure. The atomic registry between the copper (111) and hexagonal graphene was observed to enhance the transport across the copper grains significantly. The influence of grain disorder at varying densities and with oxidation groups, such as oxides and hydroxides (common in coal-derived graphene), on the electronic conductivity was analyzed. We observed a migration of functional groups from the graphene interface to the copper interface, which significantly influences the structure of the interface and electronic properties. Strong reactivity between copper and oxygen was observed, leading to the formation of metal oxides in the microstructure of the material, which can be observed in experimentally extruded copper-graphene composites. These oxidation groups induce localized states below the Fermi level that act as carrier trapping sites, as evidenced by the accumulation of the Bader charge of $\approx$ 1 electron and limiting the charge transfer to the carbon atoms from the copper interface. Additionally, the connecting atoms common to pentagonal and heptagonal rings in graphene deteriorate the in-plane conductivity. The undulating structure, due to non-hexagonal puckering, affects the overlap between the graphene $\pi$-orbitals and copper's electron sea, leading to partial interaction between copper and graphene along the transverse direction. We show this with the SPC plots.\\

This work presents one of the first simulations of the potential of coal-derived graphene in the development of high-performance metal-coal conductors for motor applications, thereby guiding engineers in optimizing the various factors and physical conditions during composite extrusion. For completeness, realistic calculations of conductivity at higher temperatures require phonon scattering; this can be addressed within an adiabatic approximation (thermal MD simulations on the canonical ensemble and by averaging the Kubo–Greenwood formula). The temperature dependence of electronic conductivity in composites will be detailed in the subsequent work.

\section*{Funding}
\noindent This material is based upon work supported by the Department of Energy under Award Number DE-FE0032277 and the US National Science Foundation (NSF) under award MRI 2320493. K.N. and D.A.D. used the computational resources at the Pittsburgh Supercomputing Center (Bridges2 Regular Memory) through allocation MAT240030, from the Advanced Cyberinfrastructure Coordination Ecosystem: Services \& Support (ACCESS) program, funded by the US NSF grants: 2138259, 2138286, 2138307, 2137603, and 2138296.

\section*{Disclaimer}
\noindent This report was prepared as an account of work sponsored by an agency of the United States Government. Neither the United States Government nor any agency thereof, nor any of their employees, makes any warranty, express or implied, or assumes any legal liability or responsibility for the accuracy, completeness, or usefulness of any information, apparatus, product, or process disclosed, or represents that its use would not infringe privately owned rights. Reference herein to any specific commercial product, process, or service by trade name, trademark, manufacturer, or otherwise does not necessarily constitute or imply its endorsement, recommendation, or favoring by the United States Government or any agency thereof. The views and opinions of authors expressed herein do not necessarily state or reflect those of the United States Government or any agency thereof.

\section*{Data Availability Statement}

The data that support the findings of this study are available within the article. \\

\noindent \textbf{References}

\nocite{*}
\bibliography{CMC}% Produces the bibliography via BibTeX.

\appendix

\section{Space-projected Conductivity (SPC) }\label{Theory_KGF_SPC}

In the single-particle approximation, the electrical conductivity is determined by averaging the diagonal elements of the conductivity tensor ($\sigma_{\alpha\alpha}$), with $\alpha$ representing the Cartesian coordinate indices $(x, y, z)$, for any frequency $\omega$.
   
\begin{multline}\label{kgf}
    \sigma(\omega) = \frac{2\pi e^2}{3m^2\Omega \omega} {\sum_\textit{\textbf{k}} w_\textbf{k} \sum_{i,j} \sum_{\alpha} [ f(\epsilon_{i,\textit{\textbf{k}}}) - f(\epsilon_{j,\textit{\textbf{k}}})]} \\
    |\langle\psi_{j,\textit{\textbf{k}}}| \textit{\textbf{P}}^{\alpha}|\psi_{i,\textit{\textbf{k}}}\rangle|^2 \delta(\epsilon_{j,\textit{\textbf{k}}} - \epsilon_{i,\textit{\textbf{k}}} - \hbar \omega)
\end{multline} \\

where, $e$ and $m$ represent the electronic charge and mass, respectively, $\Omega$ is the volume of the supercell, and $w_{\textbf{\textit{k}}}$ are the integration weight factors for \textbf{\textit{k}}-points. The single-particle Kohn-Sham states, $\psi_{i,\textbf{\textit{k}}}$, are associated with energies $\epsilon_{i,\textbf{\textit{k}}}$ for band index $i$ and Bloch vector \textbf{\textit{k}}. The momentum operator along the $\alpha$ direction is denoted by $\textbf{\textit{P}}^{\alpha}$, and $f(\epsilon_{i,\textbf{\textit{k}}})$ represents the Fermi-Dirac distribution weight. 

The method of SPC \cite{subedi_pssb} follows from Equation \ref{kgf}, by defining:

\begin{equation}
g_{i,j,\textit{\textbf{k}}} = \frac{2\pi e^2}{3m^2\Omega \omega} \left [ f(\epsilon_{i,\textit{\textbf{k}}}) - f(\epsilon_{j,\textit{\textbf{k}}}) \right ]  \delta \left(\epsilon_{j,\textit{\textbf{k}}} - \epsilon_{i,\textit{\textbf{k}}} - \hbar \omega \right )
\end{equation}

\noindent Next, a complex-valued function: $\xi_{ij\textit{\textbf{k}}}^{\alpha}(x) =  \psi_{j,\textit{\textbf{k}}}^*(x)\textit{\textbf{P}}^{\alpha} \psi_{i,\textit{\textbf{k}}}(x)$ on a discrete real-space grid (\textit{x}) uniformly spaced in a 3D supercell with spacing \textit{h} is defined. Equation \ref{kgf} is rewritten as 
   \begin{equation}\label{kgf3}
    \sigma \approx h^6 {\sum_\textit{\textbf{k}} w_\textit{\textbf{k}} \sum_{i,j,\alpha}} \sum_{x,x'} g_{i,j,\textit{\textbf{k}}} \left [\xi_{ij\textit{\textbf{k}}}^{\alpha}(x) \right ]\left [\xi_{ij\textit{\textbf{k}}}^{\alpha}(x') \right ]^*
    \end{equation}
    
\noindent where the conduction matrix ($\Gamma$) is given as

 \begin{equation}\label{kgf4}
    \Gamma(x,x') =  h^6 {\sum_\textit{\textbf{k}} w_\textit{\textbf{k}} \sum_{i,j,\alpha}} g_{i,j,\textit{\textbf{k}}} \left [\xi_{ij\textit{\textbf{k}}}^{\alpha}(x) \right ]\left [\xi_{ij\textit{\textbf{k}}}^{\alpha}(x') \right ]^*
    \end{equation}
    
\noindent The matrix elements of $\Gamma$ have the dimension of conductivity, and summation over all grid points recovers the KGF conductivity \ref{kgf} in a limit as $h \rightarrow 0$. Summing out $x^\prime$ gives the SPC $(\zeta)$ at real-space grid points $x$.

 \begin{equation}\label{spc}
    \zeta(x) = \left| \sum_{x^\prime} \Gamma(x,x^\prime) \right |
    \end{equation}
The modulus is taken to ensure a real value for the scalar field $\zeta$. References \cite{Subedi_2022,subedi_pssb, Prasai_2018, Subedi_2019} detail the application of space-projected conductivity for an atomistic study of the transport properties in a wide range of materials--semiconductors, disordered systems, and composites. \\
\\

\end{document}